\documentclass[sigplan,10pt]{acmart}
\usepackage{subcaption}
\usepackage{placeins}
\usepackage{float}
\renewcommand\footnotetextcopyrightpermission[1]{}
\title{Poisoning Learned Index Structures: Static and Dynamic Adversarial Attacks on ALEX}

\author{Allen Jue}
\affiliation{
  \institution{The University of Texas at Austin}
  \city{Austin}
  \state{Texas}
  \country{USA}
}
\email{mrallenjue@utexas.edu}
\begin{document}

\begin{abstract}
Learned index structures achieve high performance by modeling the cumulative distribution function (CDF) of keys, but this reliance on data distributions introduces potential vulnerability to adversarial manipulation. Prior work has explored both static data poisoning and dynamic algorithmic complexity attacks (ACA), though evaluations are typically limited in scale or consider only one threat model. We present a systematic study of both attack paradigms on ALEX, a state-of-the-art dynamic learned index, under a unified and reproducible framework. 

Our evaluation scales to realistic workloads with up to 200K adversarial inserts and includes multiple SOSD datasets with diverse key distributions, as well as a real-key baseline to isolate adversarial effects. Our results show a clear separation between threat models. Static poisoning has minimal impact on lookup performance in ALEX under bulk-loaded settings, while dynamic ACA induces substantial degradation, with up to 2–2.8$\times$ slowdown in lookup throughput. However, attack effectiveness is highly dataset-dependent: dense key distributions limit adversarial leverage due to duplicate-heavy insertions and ALEX’s localized structure. 

We highlight key evaluation considerations, including the need for control workloads and the mismatch between localized structural damage and global query metrics. These results show that robustness in learned indexes depends critically on the interaction between threat model, data distribution, and evaluation methodology.
\end{abstract}

\maketitle
\pagestyle{plain}
\pagenumbering{arabic}

\section{Introduction}

Modern database systems rely on a core set of data structures to support efficient query processing. Structures such as B-trees provide ordered range scans and point lookups with $O(\log n)$ worst-case guarantees. These structures are highly optimized and well understood: their performance bounds are tight, their behavior is predictable, and they make no assumptions about the distribution of the underlying data.

This generality is also a limitation. A B-tree indexing sequential user IDs behaves identically to one indexing highly skewed geographic coordinates. The structure is oblivious to the statistical properties of the data it stores and does not exploit patterns in key distributions. As a result, every lookup incurs the same asymptotic cost regardless of whether the data is uniform, clustered, or heavily skewed.

Learned index structures address this limitation by replacing parts of the index with learned models. First introduced in \citet{46518}, these systems treat indexing as a prediction problem: given a key $k$, predict its position in the sorted dataset by approximating the cumulative distribution function (CDF) of the data. Simple distributions can be modeled with linear functions, while more complex distributions require hierarchical models that partition the key space. Systems such as ALEX~\cite{ALEX}, PGM-Index~\cite{PGM-index}, and FINEdex~\cite{Finedex} demonstrate that CDF-aware indexes can achieve substantial improvements in lookup latency and memory usage, particularly on structured real-world datasets.

However, this dependence on data distribution introduces a vulnerability. Learned indexes are trained on the statistical properties of their input and updated as new keys are inserted, making their performance sensitive to changes in the underlying distribution. An adversary with write access can exploit this by inserting carefully chosen keys that distort the learned CDF, degrading model accuracy and increasing lookup cost. This class of manipulation is known as a data poisoning attack. Prior work has shown that such attacks on recursive model indexes (RMI) can increase prediction error by up to 24--70$\times$, significantly affecting performance~\cite{ACM_Data_poisoning}. In contrast, traditional structures such as B-trees and hash tables remain unaffected due to their distribution-agnostic design.

Subsequent work has explored the robustness of learned indexes under static adversarial conditions. For example, evaluations of ALEX under poisoning scenarios suggest limited sensitivity to certain attacks once realistic workloads are applied~\cite{bachfischer2022testingrobustnesslearnedindex}. Separately, algorithmic complexity attacks (ACA) on dynamic learned indexes demonstrate that adversarial query patterns can trigger worst-case structural behavior~\cite{yang2024algorithmiccomplexityattacksdynamic}. These results indicate that learned indexes are sensitive not only to training-time data, but also to runtime workload dynamics.

In this work, we study the robustness of learned indexes under both static and dynamic adversarial workloads using ALEX as a representative learned index and Abseil’s B-tree as a classical baseline. We find that static poisoning has limited persistent impact once the system adapts to query execution, whereas dynamic adversaries that interact with the index during runtime can induce substantial and sustained degradation.

We evaluate two adversarial settings. First, a static poisoning attacker constructs adversarial datasets drawn from the Search on Sorted Data (SOSD) benchmark~\cite{sosd-neurips}, with the objective of maximizing CDF approximation error at build time. Second, we implement an algorithmic complexity attack (ACA) based on adaptive insert-and-query cycles, instantiated in both black-box and white-box variants.

Across multiple real-world datasets (Books, Facebook, Lognormal, and Wiki TS), dynamic ACA attacks consistently produce significantly greater degradation than static poisoning. In particular, ACA increases per-lookup latency on real keys by up to 2--2.8$\times$ at peak, measured in nanoseconds per operation. This degradation arises from structural drift induced by adversarial insertions rather than initial model error, whereas the effects of static poisoning diminish as the system adapts.

These results suggest that robustness evaluations of learned indexes must account for interactive adversarial workloads rather than relying solely on static dataset manipulation. Performance stability depends not only on model accuracy at construction time, but also on the interaction between workload dynamics, index maintenance, and locality-sensitive updates.

\paragraph{Contributions.}
This paper makes the following contributions:
\begin{itemize}
    \item We evaluate the robustness of \textsc{ALEX} under both static poisoning and dynamic adversarial workloads, using Abseil’s B-tree as a classical index baseline.
    
    \item We empirically demonstrate that dynamic adversaries induce substantially greater and more persistent performance degradation than static poisoning across multiple real-world datasets.
\end{itemize}




\section{Related Work}

\subsection{Learned Index Structures}

Learned index structures model indexing as a prediction problem by approximating the cumulative distribution function (CDF) of keys~\cite{46518}. Early designs such as the recursive model index (RMI) demonstrate that learned models can replace traditional tree-based structures for point and range queries. Subsequent systems improve robustness, efficiency, and support for dynamic updates.

ALEX~\cite{ALEX} is an updatable in-memory learned index that combines an RMI-style hierarchy of linear models with gapped arrays. It adapts to insertions by expanding or splitting nodes based on a cost model, enabling efficient performance under evolving workloads. The PGM-index~\cite{PGM-index} uses piecewise linear models with provable worst-case error bounds and supports dynamic updates via buffering and merging. Other systems such as FINEdex~\cite{Finedex} further explore scalability and concurrency in learned indexing. Surveys of the area highlight significant performance improvements while identifying robustness and security as open challenges~\cite{learned_survey}.

\subsection{Poisoning Attacks on Learned Indexes}

The reliance of learned indexes on data distributions introduces vulnerability to data poisoning attacks. Kornaropoulos et al.~\cite{ACM_Data_poisoning} present the first systematic study of such attacks, showing that carefully crafted modifications to the training data can significantly increase model error and degrade performance across a variety of learned index designs.

Subsequent work has examined the robustness of learned indexes under static adversarial settings. Bachfischer et al.~\cite{bachfischer2022testingrobustnesslearnedindex} evaluate multiple learned index structures, including ALEX, and observe limited sensitivity to certain poisoning strategies under their experimental conditions. However, their evaluation focuses on relatively small-scale and synthetic datasets, leaving open how these attacks behave in larger or more dynamic environments.

Recent theoretical work further analyzes the behavior of poisoning attacks on CDF-based models. Sato et al.~\cite{sato2026mathematicalfoundationspoisoningattacks} study the effects of adversarial perturbations on linear regression over cumulative distributions, providing insight into how poisoning can systematically distort learned mappings.

\subsection{Algorithmic Complexity Attacks}

Algorithmic complexity attacks on learned indexes have been explored in recent work~\cite{yang2024algorithmiccomplexityattacksdynamic}. These attacks exploit worst-case structural behavior through adversarial query patterns, demonstrating that workload dynamics can significantly impact performance. Unlike poisoning attacks, which target model accuracy at training time, these approaches focus on triggering inefficient execution paths during query processing.

\section{Methodology}

\subsection{System and Experimental Setup}

We evaluate learned index robustness using \textsc{ALEX}~\cite{ALEX} as the representative learned index and Abseil’s B-tree~\cite{abseil} as a classical baseline for static experiments. ALEX is integrated into our benchmarking harness based on the SOSD framework, and core ALEX functionality, including insertion and node-splitting behavior, is left unchanged. Minor modifications are limited to portability fixes (e.g., ARM-safe instructions), while all adversarial logic is implemented externally within the benchmarking code.

Experiments are conducted on a machine with an Apple M4 Max CPU (16 cores) and 64\,GB of RAM. Static experiments use Google Benchmark~\cite{googlebenchmark} to measure lookup latency, while dynamic experiments use a custom timing harness based on \texttt{std::chrono}, as the Google Benchmark is best-suited for stationary workloads. Reported results include throughput (in millions of operations per second) and derived latency metrics.

Datasets are drawn from the SOSD benchmark suite~\cite{sosd-neurips, sosd-vldb}. Raw datasets contain up to $10^8$ keys; however, for experimental efficiency, we preprocess them by subsampling and deduplication. Static experiments operate on datasets of up to $10^5$ keys, while dynamic ACA experiments use up to $10^6$ keys with subsampling, resulting in effective dataset sizes on the order of $10^5$--$10^6$ keys depending on duplication. Queries are sampled uniformly over the set of clean keys.

\subsection{Adversarial Models}

We consider two adversarial settings: static data poisoning and dynamic adversarial workloads.

In the static setting, the adversary modifies the dataset prior to index construction. Given a clean dataset of size $N$, the adversary inserts a set of adversarial keys of size $\alpha N$, where $\alpha \in \{0.05, 0.10, 0.15, 0.20\}$. The goal is to maximize prediction error of the learned model, thereby increasing lookup cost after bulk loading.

In the dynamic setting, the adversary interacts with the index after construction through a sequence of insertions. Starting from a clean index, the adversary inserts keys in batches over multiple rounds. After each batch, lookup performance is measured on the original (clean) keys. This process continues until a total of $\alpha N$ adversarial keys have been inserted. Unlike static poisoning, this adversary influences both the data distribution and the structure of the index over time.

\subsection{Attack Implementation}

\paragraph{Static Poisoning.}
Adversarial keys are generated using a greedy objective that approximates maximization of the mean squared error (MSE) of the CDF model. At each step, candidate keys are selected from integer gaps between sorted neighboring keys, and the key that most increases model error is chosen. This process continues until the poisoning budget is reached or no valid insertion positions remain. The resulting poisoned dataset is then used to bulk-load the index.

\paragraph{Dynamic ACA}
We implement an algorithmic complexity attack (ACA) based on adaptive insert-and-query cycles. Let $B$ denote the batch size and $R = \lceil \alpha N / B \rceil$ the number of rounds. Each round consists of inserting a batch of $B$ adversarial keys followed by a lookup phase over clean keys.

We consider two variants of ACA:

\emph{Black-box ACA} generates adversarial keys without access to internal index state. Keys are sampled from the global key range using randomized strategies, without using feedback from observed latency.

\emph{White-box ACA} uses access to internal index structure to guide key selection. At each round, the adversary identifies the largest data node in the index and locates the longest contiguous occupied segment within that node. New keys are inserted immediately adjacent to this segment, promoting local density and triggering structural changes such as node expansion or splitting. This process is repeated adaptively as the index evolves.

In both variants, adversarial key selection does not directly optimize measured latency; instead, performance degradation emerges from the cumulative structural effects of insertions.

\subsection{Evaluation Protocol}

Performance is measured on lookup operations over clean keys. For static experiments, we report lookup latency using Google Benchmark and normalize results relative to a clean baseline. For dynamic ACA experiments, we report throughput in millions of operations per second (Mops/s) and compute a slowdown ratio relative to the clean index. Each experiment is repeated multiple times (typically 5), and we report mean and standard deviation.

In ACA experiments, each run begins with a freshly constructed index built from clean data. A warm-up phase precedes timing measurements to ensure stable results. Insertions and lookups are separated at the batch level: each round applies insertions first, followed by a lookup-only measurement phase.

For static experiments, the index is rebuilt from the corresponding clean or poisoned dataset before each benchmark run. B-tree comparisons are included only in the static setting to contrast learned and traditional index behavior under poisoning.

\section{Results}

\subsection{Static Poisoning}

\begin{figure*}[t]
    \centering
    \includegraphics[width=\textwidth]{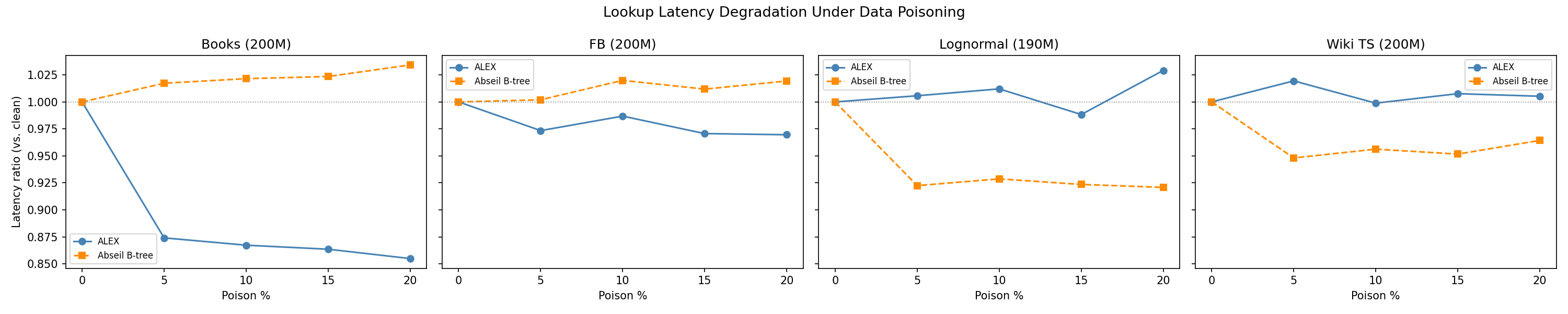}
    \caption{Lookup slowdown under static poisoning as a function of poisoning budget. Across all datasets, degradation remains minimal and non-monotonic, with worst-case slowdown below $1.03\times$.}
    \label{fig:static_degradation}
\end{figure*}

\begin{figure*}[t]
    \centering
    \includegraphics[width=\textwidth]{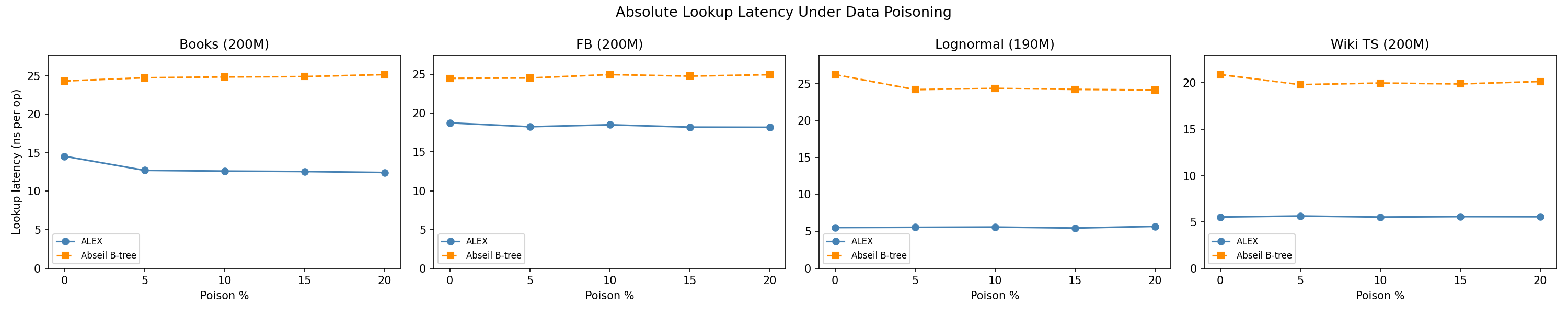}
    \caption{Absolute lookup performance under static poisoning. Variations across poisoning levels are small and largely attributable to measurement noise rather than systematic degradation.}
    \label{fig:static_absolute}
\end{figure*}

We first evaluate traditional data poisoning attacks applied prior to index construction. Figure~\ref{fig:static_degradation} shows slowdown as a function of poisoning budget across datasets.

Overall, static poisoning has limited impact on lookup performance. Across all datasets and poisoning levels, the observed slowdown ranges from $0.855\times$ to $1.029\times$, with the worst case occurring on the lognormal dataset at the highest poisoning level. The curves are largely flat and non-monotonic, indicating that variation is dominated by noise rather than systematic degradation.

These results indicate that static poisoning is insufficient to produce meaningful lookup slowdown. Even when adversarial keys are chosen to maximize model error, the resulting performance impact remains negligible. This behavior is consistent with recent theoretical analyses of CDF-based poisoning, which show that optimal adversarial keys tend to concentrate near distribution endpoints or regression intersection points rather than spreading broadly across the key space~\cite{sato2026mathematicalfoundationspoisoningattacks}. Such concentration limits their ability to induce widespread structural degradation in learned indexes.

\subsection{Dynamic Adversarial Workloads}

\begin{figure*}[t]
    \centering
    \begin{subfigure}{1.0\textwidth}
        \centering
        \includegraphics[width=\linewidth]{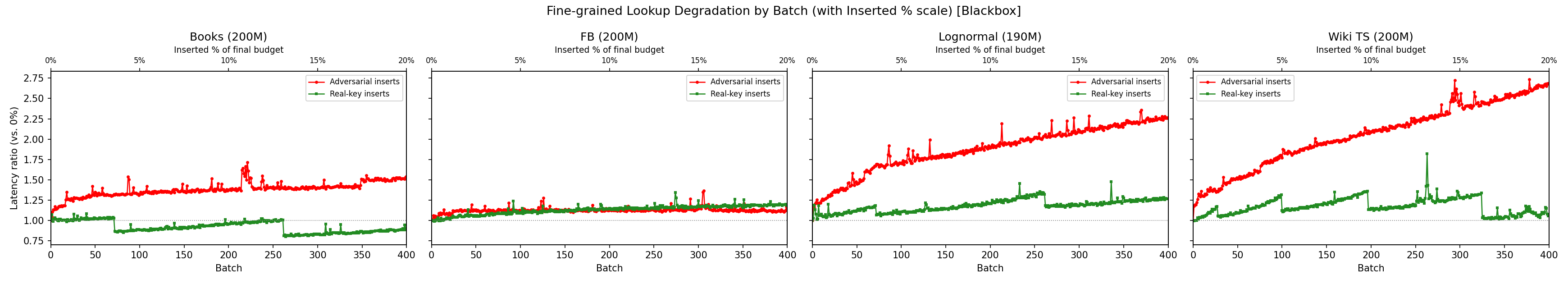}
        \caption{Black-box ACA}
    \end{subfigure}

    \vspace{0.5em}

    \begin{subfigure}{1.0\textwidth}
        \centering
        \includegraphics[width=\linewidth]{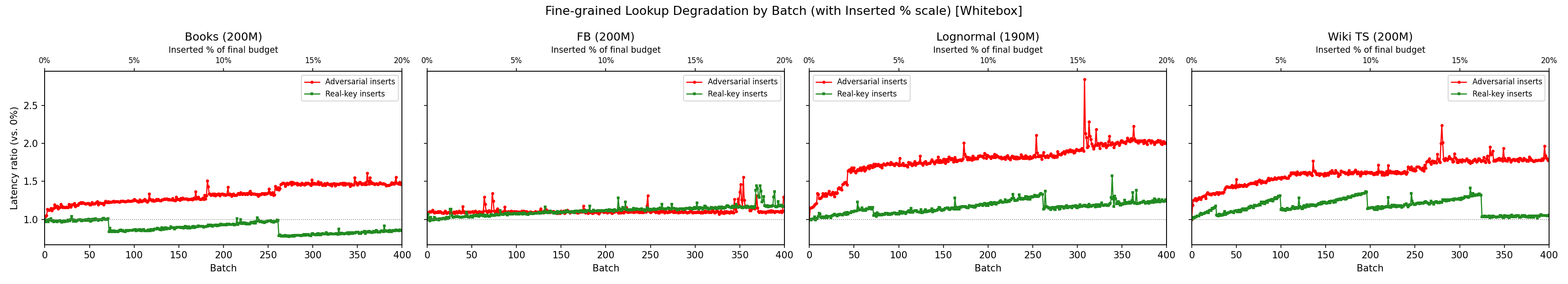}
        \caption{White-box ACA}
    \end{subfigure}

    \caption{Fine-grained slowdown over insertion rounds for black-box and white-box ACA. Both attacks induce substantial degradation, with peak slowdowns up to $2.7\times$--$2.8\times$. Jagged trajectories reflect discrete structural events such as node splits and retraining, as well as per-batch variability in insertion locality.}
    \label{fig:aca_finegrained}
\end{figure*}

We next evaluate algorithmic complexity attacks (ACA), where adversarial keys are inserted in batches and performance is measured after each round.

Dynamic adversaries induce substantially larger degradation than static poisoning. Across datasets, ACA achieves peak slowdowns between $2.7\times$ and $2.8\times$, with the strongest effects on lognormal and Wiki TS. Slowdown increases as insertions accumulate and often stabilizes at an elevated plateau.

\subsection{Black-box vs.\ White-box Attacks}

We compare black-box and white-box variants of ACA to evaluate the role of structural knowledge. Peak degradation is slightly higher for white-box attacks, but this advantage is not consistent across datasets.

At the end of the attack sequence, black-box attacks often match or exceed white-box degradation. These results indicate that explicit structural knowledge is not required to induce substantial slowdown; randomized insertion strategies are often sufficient.

\subsection{Dataset Dependence}

\begin{figure}[t]
    \centering
    \includegraphics[width=1.0\linewidth]{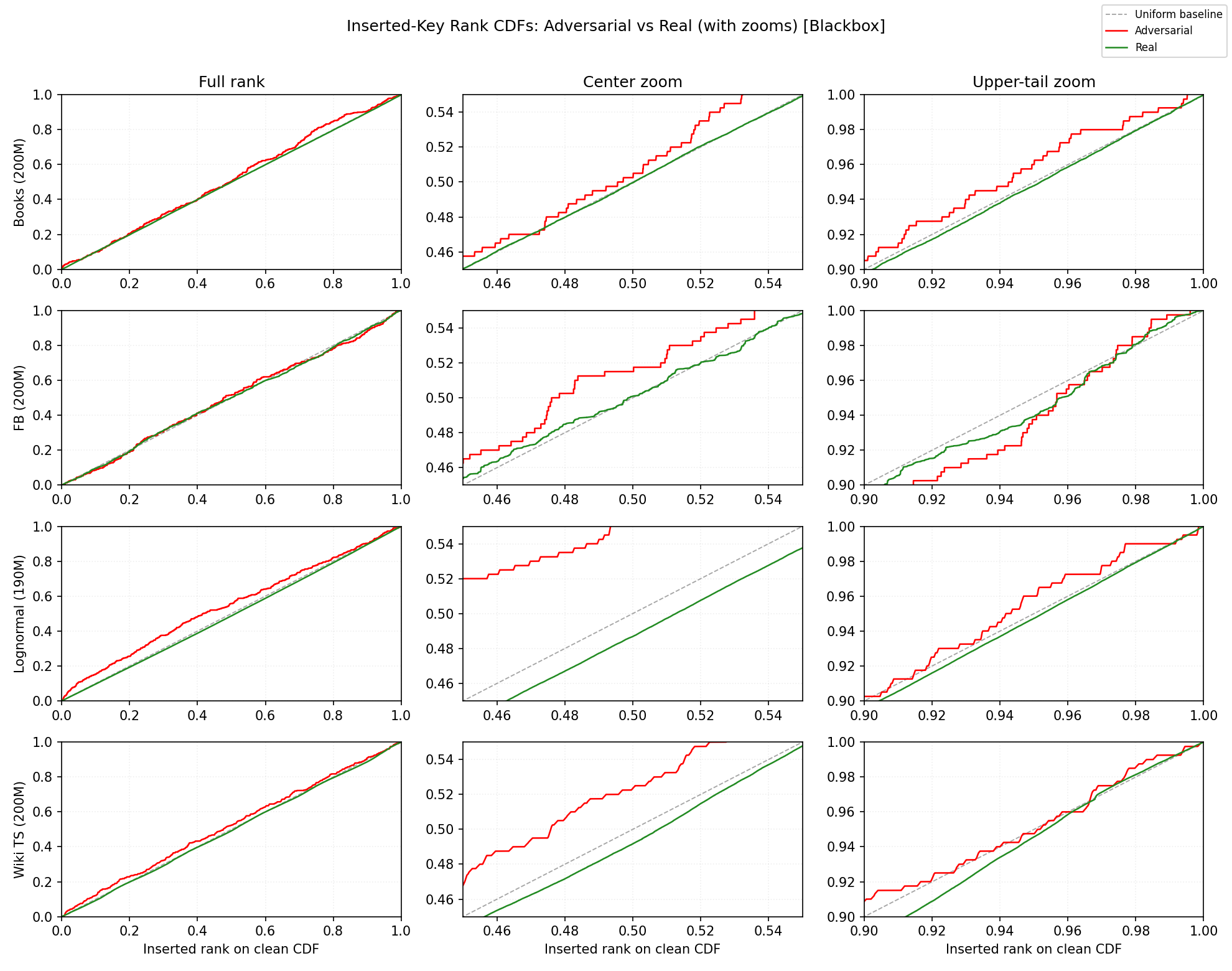}
    \caption{Empirical CDF of inserted-key positions in clean-key bins. Datasets where inserted keys occupy a larger fraction of clean-rank bins exhibit higher degradation.}
    \label{fig:rank_cdf}
\end{figure}

Attack effectiveness varies significantly across datasets. In particular, the Facebook dataset consistently exhibits the weakest degradation.

Figure~\ref{fig:rank_cdf} plots each inserted key's position in the clean-key order. On lognormal and Wiki TS, inserted keys occupy a large fraction of clean-rank bins, spreading structural pressure broadly across the index. On Facebook, inserted keys fall into a narrow band of clean-rank bins — despite appearing spread in raw value space — restricting damage to a small region of the index.

Additional distribution visualizations are provided in Appendix~\ref{sec:appendix_distributions}.

\subsection{Summary of Findings}

Static poisoning produces minimal degradation, while dynamic adversaries induce substantial and sustained slowdown. These results show that robustness depends on the interaction between adversarial insertions, workload dynamics, and the locality-sensitive structure of the index.

\section{Discussion}

Static poisoning and dynamic ACA stress fundamentally different failure modes. Static attacks corrupt the training distribution at build time; dynamic ACA exploits the update path at runtime. The consistent gap in our results — negligible slowdown under static poisoning versus up to $2.8\times$ under ACA — reflects this architectural distinction rather than a difference in attack intensity.

ACA effectiveness stems from interaction with ALEX's maintenance logic. Each adversarial batch modifies local node density, triggering expansion and splitting events that reshape the index hierarchy. The plateau-forming, jagged trajectories in Figure~\ref{fig:aca_finegrained} reflect discrete structural events rather than a smooth function of insert count — a qualitatively different failure mode from model error that calls for different mitigations. Notably, white-box ACA does not consistently outperform black-box at end-state, implying that learned indexes are vulnerable to relatively unsophisticated adaptive adversaries.

Attack effectiveness is further governed by how broadly inserted keys are distributed relative to the clean data — specifically, what fraction of clean-rank bins they occupy. Lognormal and Wiki TS see insertions spread across most of the clean-rank domain, producing widespread structural disruption. Facebook's insertions, despite appearing broadly spread in raw value space, fall into a narrow band of clean-rank bins, restricting impact to a small index region. Raw value-space plots alone cannot explain this variation; plotting insertion positions in the clean-key order is necessary.

An important boundary condition is what happens when the attacker lacks write access entirely. Even without adversarial insertions, naturally skewed or heavy-tailed key distributions — such as those arising from organic workload growth or data migration — can produce similarly narrow clean-rank bin coverage that resembles adversarial conditions. This suggests that some ``degradation'' attributed to attacks in prior work may in fact reflect distribution sensitivity inherent to the learned model, not adversarial intent. Distinguishing genuine attack signal from distributional stress requires the real-key control baseline we include here.

\section{Limitations}

Our evaluation has several limitations. First, measurement phases separate insertions and lookups temporally, so mixed insert-plus-lookup service cost under realistic workloads is not captured. Second, all dynamic results are specific to ALEX; generalization to other learned indexes requires direct replication. Third, absolute throughput figures are platform-specific and should not be compared across hardware configurations — relative slowdown ratios are more robust. Finally, this work characterizes attacks only; mitigations such as CDF-influence-based outlier detection or structural monitoring heuristics are not evaluated.

\section{Conclusion}

We evaluated the robustness of ALEX under static data poisoning and dynamic algorithmic complexity attacks across four SOSD datasets. Static poisoning produces at most $1.03\times$ slowdown across all datasets and budgets, while dynamic ACA induces peak slowdowns of up to $2.8\times$ with sustained plateaus above baseline.

This separation reflects distinct failure modes. Static poisoning distorts CDF accuracy at build time, but ALEX's adaptive architecture absorbs this distortion during execution. Dynamic ACA accumulates structural drift through repeated insert-triggered maintenance events, producing degradation that grows with adversarial pressure rather than diminishing with adaptation. The fraction of clean-rank bins occupied by inserted keys 
predicts global impact better than value-space distribution. 
Black-box attacks match white-box at end-state, indicating 
that structural knowledge provides only marginal advantage.

These results establish that learned index robustness is a dynamic property. Future work should adopt interactive adversarial workloads, include real-key controls to separate adversarial effects from distributional sensitivity, and evaluate defenses against insert-path ACA.
\bibliographystyle{ACM-Reference-Format}
\bibliography{references}

@inproceedings{46518,title	= {The Case for Learned Index Structures},author	= {Tim Kraska and Alex Beutel and Ed H. Chi and Jeff Dean and Neoklis Polyzotis},year	= {2018},URL	= {https://arxiv.org/abs/1712.01208}}

@inproceedings{ALEX, series={SIGMOD/PODS ’20},
   title={ALEX: An Updatable Adaptive Learned Index},
   url={http://dx.doi.org/10.1145/3318464.3389711},
   DOI={10.1145/3318464.3389711},
   booktitle={Proceedings of the 2020 ACM SIGMOD International Conference on Management of Data},
   publisher={ACM},
   author={Ding, Jialin and Minhas, Umar Farooq and Yu, Jia and Wang, Chi and Do, Jaeyoung and Li, Yinan and Zhang, Hantian and Chandramouli, Badrish and Gehrke, Johannes and Kossmann, Donald and Lomet, David and Kraska, Tim},
   year={2020},
   month=may, pages={969–984},
   collection={SIGMOD/PODS ’20} }

@article{PGM-index,
   title={The PGM-index: a fully-dynamic compressed learned index with provable worst-case bounds},
   volume={13},
   ISSN={2150-8097},
   url={http://dx.doi.org/10.14778/3389133.3389135},
   DOI={10.14778/3389133.3389135},
   number={8},
   journal={Proceedings of the VLDB Endowment},
   publisher={Association for Computing Machinery (ACM)},
   author={Ferragina, Paolo and Vinciguerra, Giorgio},
   year={2020},
   month=apr, pages={1162–1175} }

@article{Finedex,
author = {Li, Pengfei and Hua, Yu and Jia, Jingnan and Zuo, Pengfei},
title = {FINEdex: a fine-grained learned index scheme for scalable and concurrent memory systems},
year = {2021},
issue_date = {October 2021},
publisher = {VLDB Endowment},
volume = {15},
number = {2},
issn = {2150-8097},
url = {https://doi.org/10.14778/3489496.3489512},
doi = {10.14778/3489496.3489512},
journal = {Proc. VLDB Endow.},
month = oct,
pages = {321–334},
numpages = {14}
}

@inproceedings{ACM_Data_poisoning,
author = {Kornaropoulos, Evgenios M. and Ren, Silei and Tamassia, Roberto},
title = {The Price of Tailoring the Index to Your Data: Poisoning Attacks on Learned Index Structures},
year = {2022},
isbn = {9781450392495},
publisher = {Association for Computing Machinery},
address = {New York, NY, USA},
url = {https://doi.org/10.1145/3514221.3517867},
doi = {10.1145/3514221.3517867},
booktitle = {Proceedings of the 2022 International Conference on Management of Data},
pages = {1331–1344},
numpages = {14},
location = {Philadelphia, PA, USA},
series = {SIGMOD '22}
}

@misc{bachfischer2022testingrobustnesslearnedindex,
      title={Testing the Robustness of Learned Index Structures}, 
      author={Matthias Bachfischer and Renata Borovica-Gajic and Benjamin I. P. Rubinstein},
      year={2022},
      eprint={2207.11575},
      archivePrefix={arXiv},
      primaryClass={cs.DB},
      url={https://arxiv.org/abs/2207.11575}, 
}

@article{learned_survey,
author = {Al-Mamun, Abdullah and Wu, Hao and He, Qiyang and Wang, Jianguo and Aref, Walid G.},
title = {A Survey of Learned Indexes for the Multi-dimensional Space},
year = {2025},
issue_date = {March 2026},
publisher = {Association for Computing Machinery},
address = {New York, NY, USA},
volume = {58},
number = {4},
issn = {0360-0300},
url = {https://doi.org/10.1145/3768575},
doi = {10.1145/3768575},
journal = {ACM Comput. Surv.},
month = oct,
articleno = {96},
numpages = {37}
}

@article{sosd-vldb,
  author    = {Ryan Marcus and
               Andreas Kipf and
               Alexander van Renen and
               Mihail Stoian and
               Sanchit Misra and
               Alfons Kemper and
               Thomas Neumann and
               Tim Kraska},
  title     = {Benchmarking Learned Indexes},
  journal   = {Proc. {VLDB} Endow.},
  volume    = {14},
  number    = {1},
  pages     = {1--13},
  year      = {2020}
}

@article{sosd-neurips,
  title={SOSD: A Benchmark for Learned Indexes},
  author={Kipf, Andreas and Marcus, Ryan and van Renen, Alexander and Stoian, Mihail and Kemper, Alfons and Kraska, Tim and Neumann, Thomas},
  journal={NeurIPS Workshop on Machine Learning for Systems},
  year={2019}
}

@misc{abseil,
  title        = {Abseil: An Open-Source Collection of C++ Libraries},
  author       = {{Google}},
  year         = {2024},
  howpublished = {\url{https://abseil.io}},
  note         = {Accessed: 2026-03-13}
}

@misc{googlebenchmark,
  title        = {Google Benchmark: A Microbenchmark Support Library},
  author       = {{Google}},
  year         = {2024},
  howpublished = {\url{https://github.com/google/benchmark}},
  note         = {Accessed: 2026-03-13}
}

@misc{yang2024algorithmiccomplexityattacksdynamic,
      title={Algorithmic Complexity Attacks on Dynamic Learned Indexes}, 
      author={Rui Yang and Evgenios M. Kornaropoulos and Yue Cheng},
      year={2024},
      eprint={2403.12433},
      archivePrefix={arXiv},
      primaryClass={cs.DB},
      url={https://arxiv.org/abs/2403.12433}, 
}

@misc{sato2026mathematicalfoundationspoisoningattacks,
      title={Mathematical Foundations of Poisoning Attacks on Linear Regression over Cumulative Distribution Functions}, 
      author={Atsuki Sato and Martin Aumüller and Yusuke Matsui},
      year={2026},
      eprint={2603.00537},
      archivePrefix={arXiv},
      primaryClass={cs.LG},
      url={https://arxiv.org/abs/2603.00537}, 
}

\appendix
\section{Additional Visualizations}
\label{sec:appendix_distributions}

Figures below provide value-space CDF plots for both attack variants and their real-key controls, complementing the clean-rank bin coverage analysis in Figure~\ref{fig:rank_cdf}. The two views can diverge: keys that appear broadly spread in raw value space may occupy only a narrow fraction of clean-rank bins, and value-space plots alone cannot distinguish these cases.

\begin{figure*}[t]
    \centering
    \includegraphics[width=\textwidth]{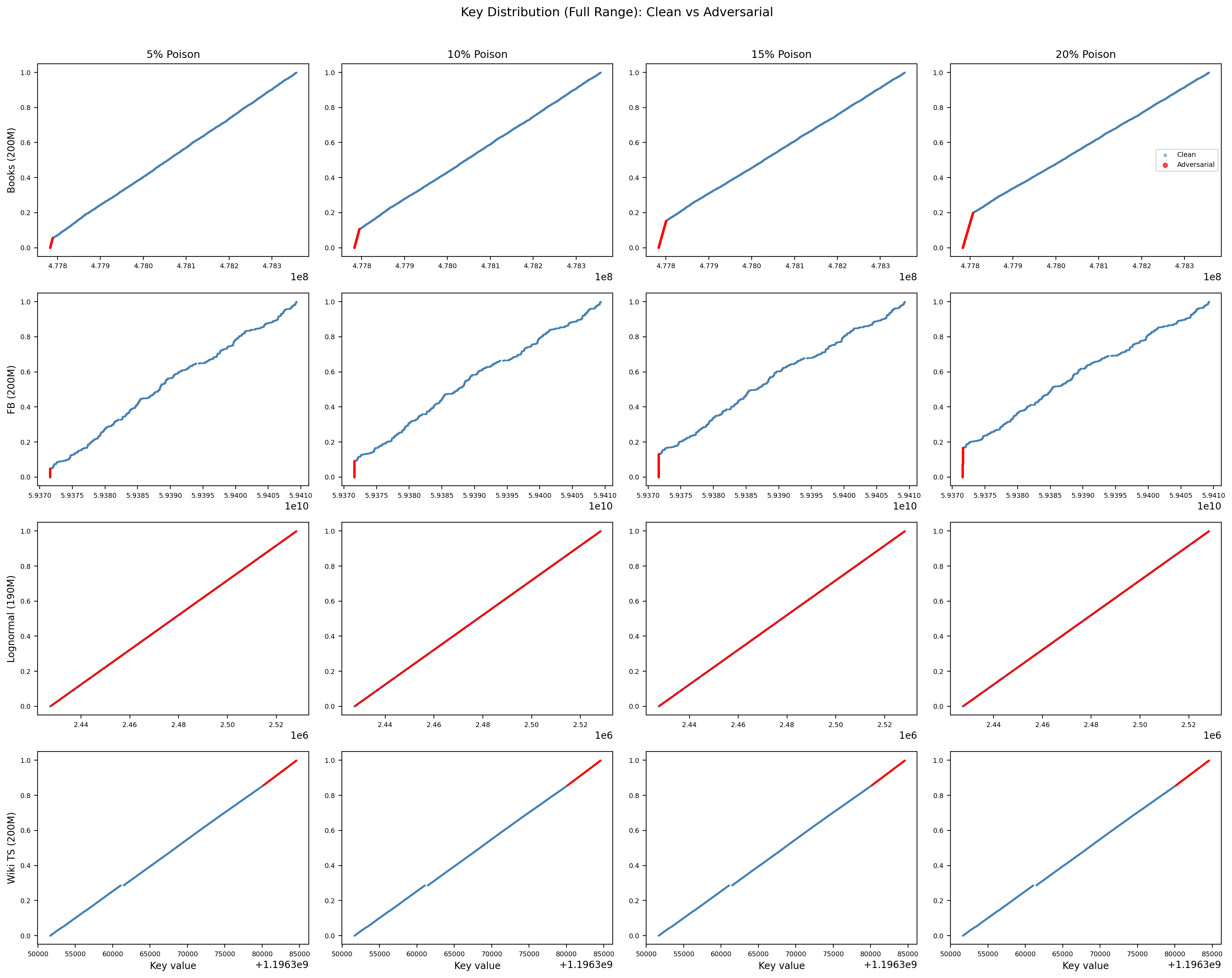}
    \caption{Static poisoning key distributions across budgets $\alpha \in \{0.05, 0.10, 0.15, 0.20\}$. Adversarial keys concentrate near distribution tails but do not induce meaningful lookup degradation.}
    \label{fig:appendix_static_dist}
\end{figure*}

\begin{figure*}[t]
    \centering
    \includegraphics[width=\textwidth]{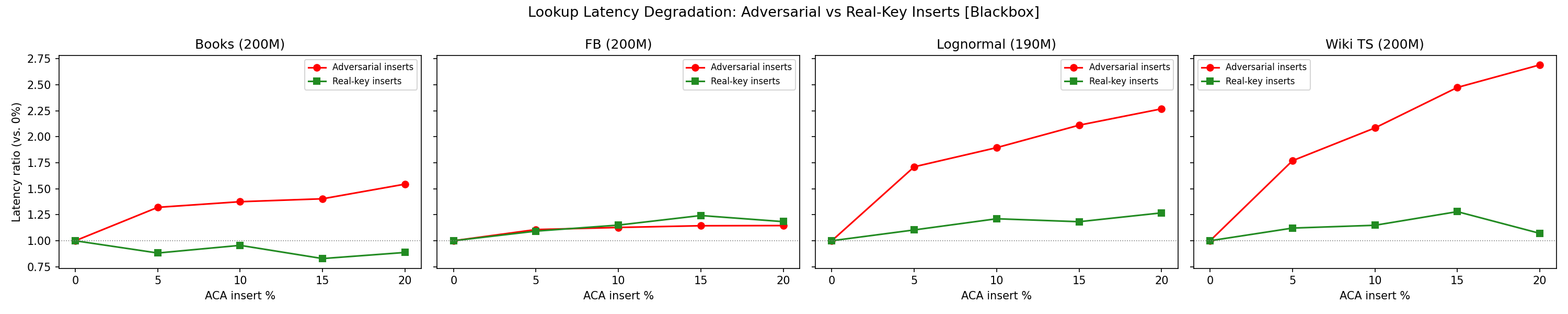}
    \caption{Black-box ACA slowdown across insertion rounds. Degradation increases over time, peaking on Lognormal and Wiki TS, while Facebook remains comparatively stable.}
    \label{fig:appendix_bb_curves}
\end{figure*}

\begin{figure*}[t]
    \centering
    \includegraphics[width=\textwidth]{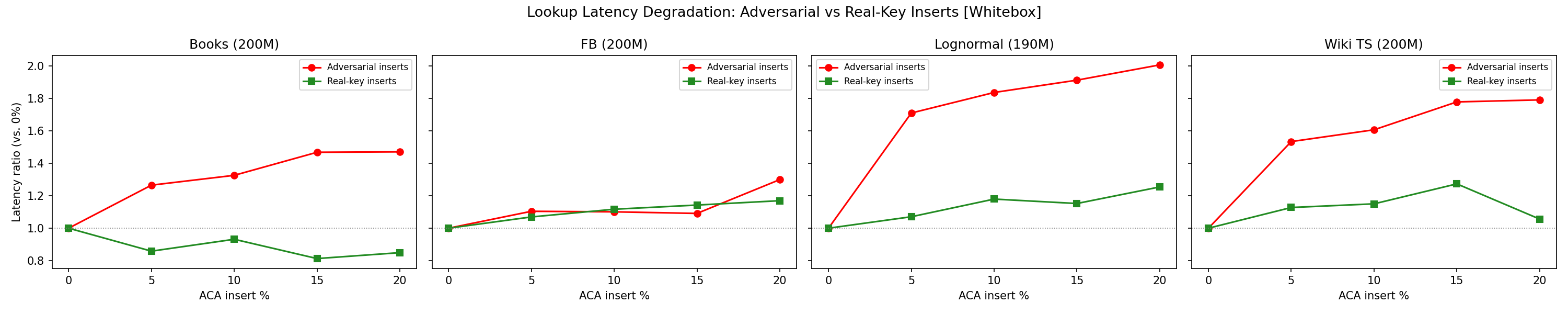}
    \caption{White-box ACA slowdown across insertion rounds. Peak degradation is comparable to black-box but not consistently higher across datasets.}
    \label{fig:appendix_wb_curves}
\end{figure*}

\begin{figure*}[t]
    \centering
    \begin{subfigure}{0.49\textwidth}
        \centering
        \includegraphics[width=\linewidth]{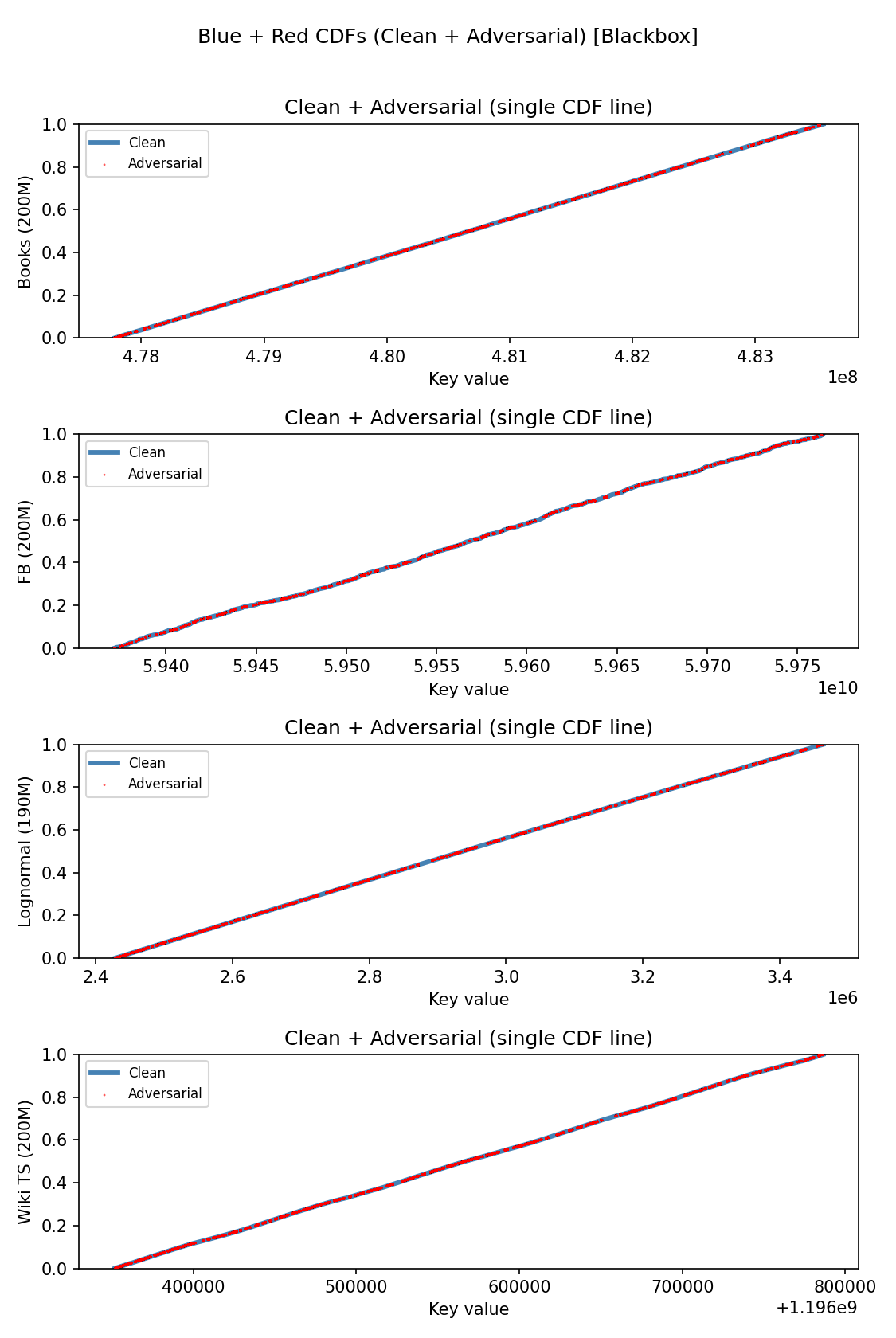}
        \caption{Black-box ACA adversarial inserts in value space. Despite apparent spread, many keys occupy a narrow band of clean-rank bins (see Figure~\ref{fig:rank_cdf}).}
        \label{fig:appendix_bb_keydist}
    \end{subfigure}
    \hfill
    \begin{subfigure}{0.49\textwidth}
        \centering
        \includegraphics[width=\linewidth]{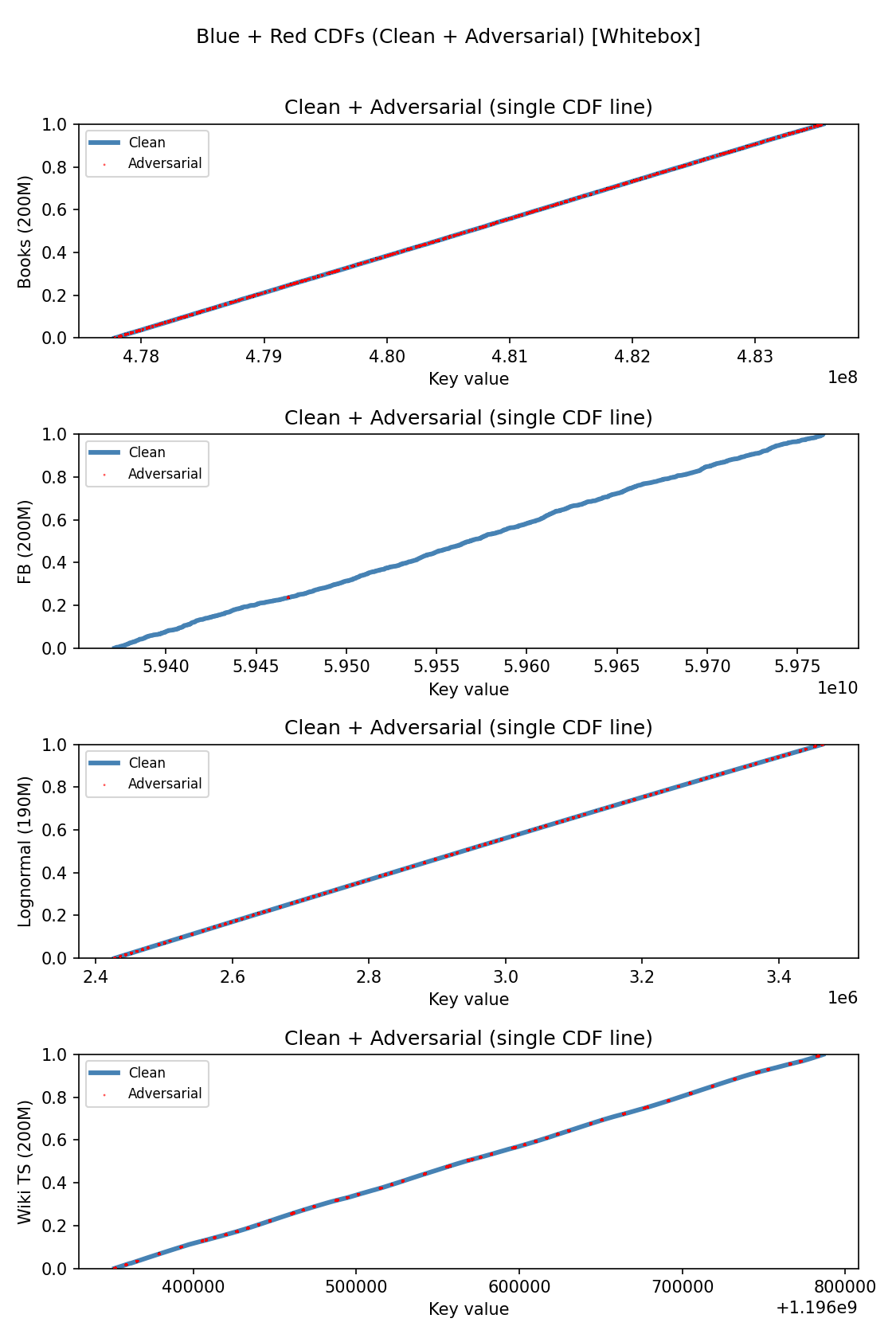}
        \caption{White-box ACA adversarial inserts in value space. Targeted insertion leads to stronger clustering in high-density regions.}
        \label{fig:appendix_wb_keydist}
    \end{subfigure}
    \caption{Adversarial key distributions in value space.}
\end{figure*}

\begin{figure*}[t]
    \centering
    \begin{subfigure}{0.49\textwidth}
        \centering
        \includegraphics[width=\linewidth]{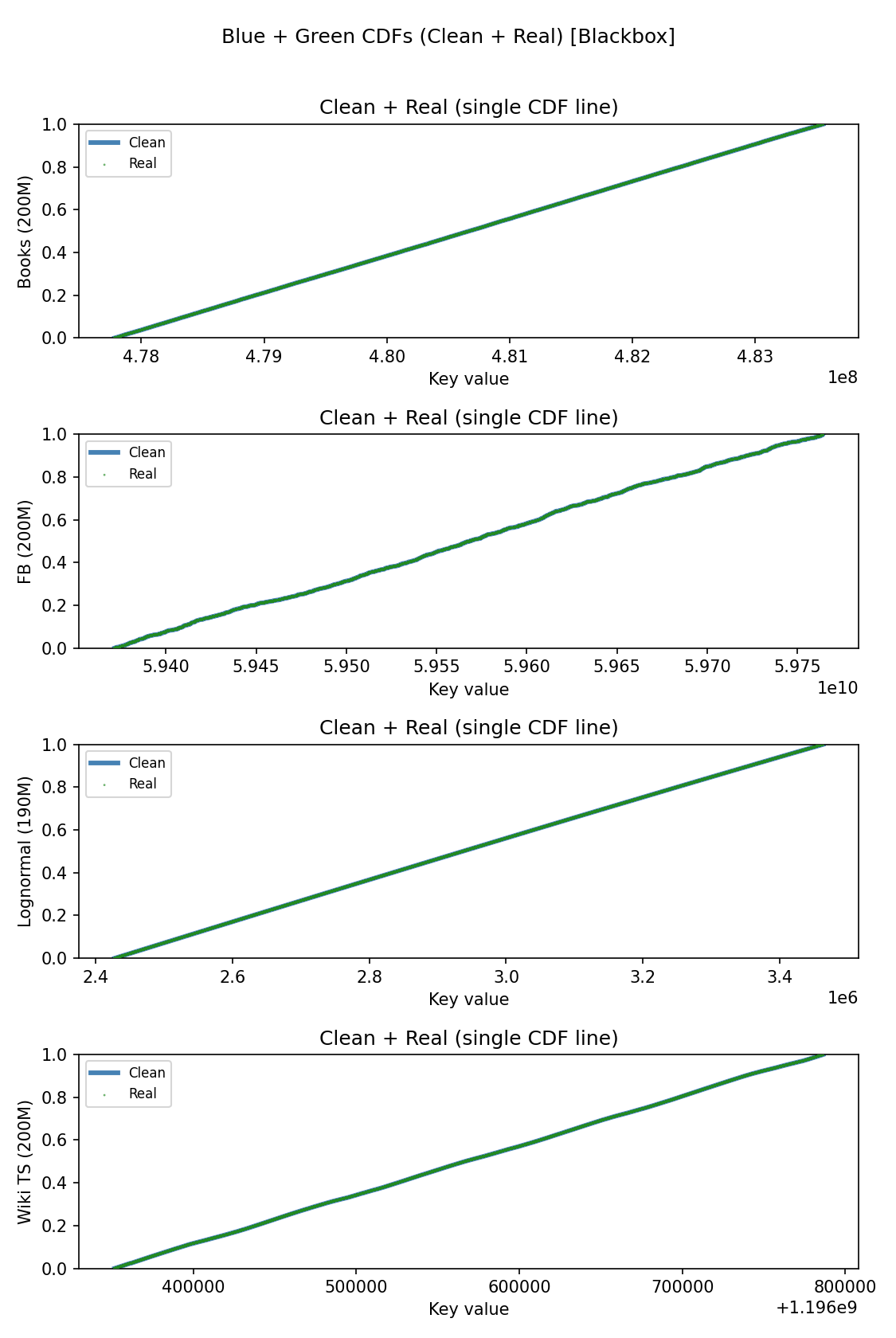}
        \caption{Black-box real-key control. Inserts follow the clean distribution closely, confirming degradation is adversarial.}
        \label{fig:appendix_bb_control}
    \end{subfigure}
    \hfill
    \begin{subfigure}{0.49\textwidth}
        \centering
        \includegraphics[width=\linewidth]{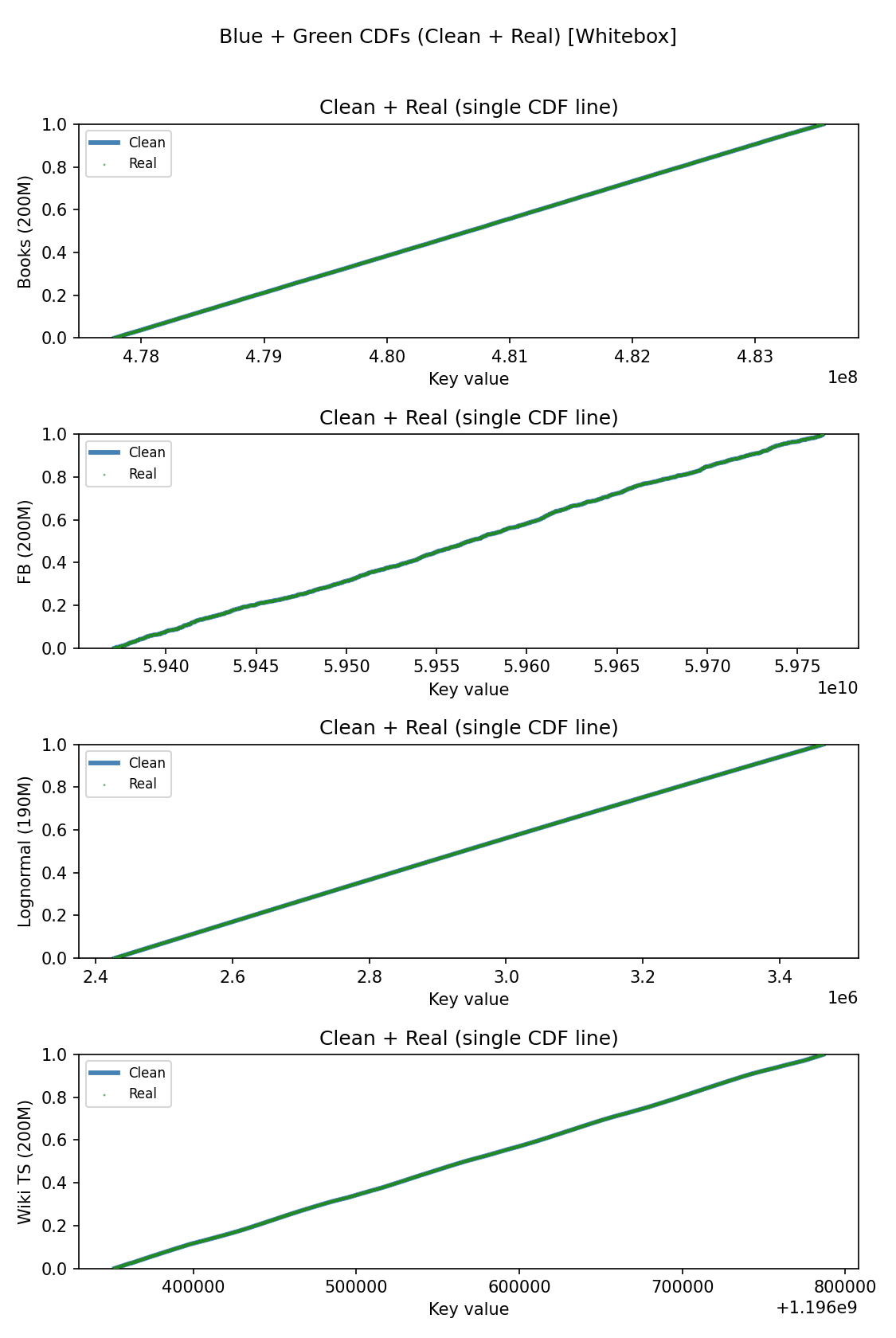}
        \caption{White-box real-key control. Closely tracks clean distribution, isolating effect of adversarial targeting.}
        \label{fig:appendix_wb_control}
    \end{subfigure}
    \caption{Real-key control distributions in value space.}
    \label{fig:appendix_controls}
\end{figure*}

\clearpage
\end{document}